\documentstyle[prl,aps]{revtex}

\input{epsf}

\epsfverbosetrue

\begin{document}
\draft
\title{Broken symmetry of row switching in 2D Josephson junction
arrays}

\author{D. Abraimov, P. Caputo, G. Filatrella\cite{addr1}, M. V. Fistul,
G. Yu. Logvenov\cite{addr2}, and A. V. Ustinov }
\address{Physikalisches Institut III, Universit\"at
Erlangen-N\"urnberg,
 D-91058 Erlangen, Germany}

\date{\today}

\wideabs{ 

\maketitle

\begin{abstract}
We present an experimental and theoretical study of row switching in
two-dimensional Josephson junction arrays. We have observed
novel dynamic states with peculiar percolative patterns
of the voltage drop inside the arrays. These states were directly
visualized using laser scanning microscopy and manifested by fine
branching in the current-voltage characteristics of the arrays.
Numerical simulations show that such percolative patterns have an
intrinsic origin and occur independently of positional disorder. We
argue that the appearance of these dynamic states is due to the
presence of various metastable superconducting states in arrays.
\end{abstract}

}

Spatiotemporal pattern formation in a system of nonlinear
oscillators is governed by mutual coupling. Examples of
such patterns are domain walls and kinks in arrays of coupled pendula,
topological spin and charge excitations in large molecules and solids,
and inhomogeneous states in many other complex systems\cite{Strogatz}. A single
driven and damped nonlinear oscillator displays two distinctly
different states, a static state and a whirling (dynamic) state.
Therefore,
a system of coupled oscillators can show various spatiotemporal patterns
with concurrently present static and dynamic states.

Arrays of Josephson junctions have attracted a lot of interest because
they serve as well-controlled laboratory
objects to study such fundamental problems as above mentioned pattern
formation, chaos, and phase locking. In the systems of coupled Josephson
junctions \cite{zant88,yu92,Ustinov,Kleiner,FilWis,Lach94,Barah98} the
two local states are the superconducting (static) state and
the resistive (dynamic) state. For underdamped junctions, these states
manifest
themselves by various branches on the current-voltage ($I$-$V$)
characteristics. The hysteretic switching between branches has been found in
stacks of Josephson tunnel junctions \cite{Ustinov}, layered
high-temperature superconductors \cite{Kleiner} and two-dimensional (2D)
Josephson junction arrays \cite{zant88,yu92,Lach94,Barah98}.

In 2D arrays, this switching effect appears in the form of {\it row
switching} when a voltage drop in array occurs on individual rows of
junctions perpendicular to the direction of the bias current. A typical
behavior found in both experiments \cite{Lach94} and in numerical
simulations \cite{yu92,Phillips:PRB-94} is that the number of rows
switched to resistive state and, even more important, their distribution
inside the array can change during sweeping the $I$-$V$ curve.
An analysis of such row switching has been carried out recently
\cite{Barah98} to predict the range of stability of the various
resistive states. However, until now only relatively simple resistive
states have been found. These states correspond to a whirling
state of {\it vertical} junctions (placed in the direction of the bias
current) and small oscillations of the Josephson phase difference of
{\it horizontal} junctions
(transverse to the bias) around their equilibrium
state \cite{Barah98}.

In this Letter, we report on the observation of new dynamic states
with peculiar percolative patterns of the voltage drop inside
{\it homogeneously} biased arrays. We have found broken symmetry in row
switching such that the dc voltage drop meanders between rows. These
states appear in dc measurements in the form of fine branching on the
$I$-$V$ curves. The states have been visualized by using the Low
Temperature Scanning Laser Microscopy (LTSLM)\cite{Sivakov}. We have also
carried out numerical simulations in which similar dynamic states are found.
The measured two-dimensional arrays consist of underdamped
Nb$/$Al-AlO$_x/$Nb Josephson junctions\cite{Hypres}. The junctions are placed at
crossing of the superconducting striplines, that are arranged in
either square lattice (with 4 junctions per elementary cell) or
triangular lattice (with 3 junctions per elementary cell). An optical
image of a 2D square array and the equivalent electrical circuit are shown
in Fig.~1.
\begin{figure}
\centering\leavevmode
\mbox{\epsfbox{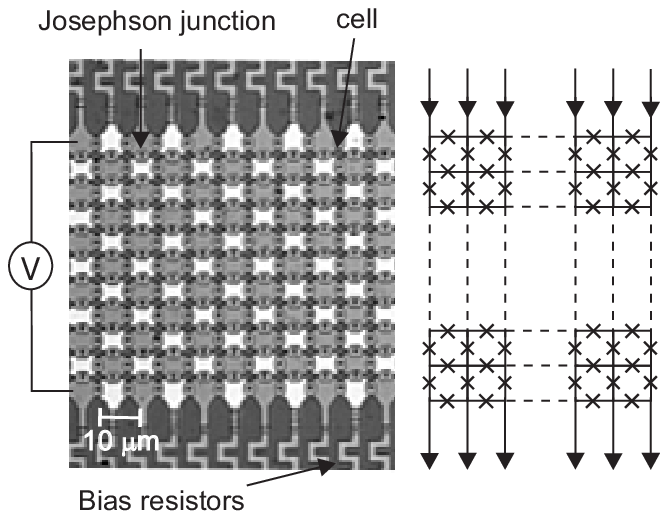}}
\caption{Optical image and sketch of $10\,\times\,10$ square array.}
\end{figure}
The junction area is 9\,$\mu$m$^2$ and their critical current
density is about $1.1$\,KA/cm$^2$ for square array, and
$0.1$\,KA/cm$^2$ for triangular array, with a typical
spread of junction parameters of about 5\%.
The square arrays consist of 10 columns by 10 rows
and have the cell area $S\,=\,38\,\mu$m$^2$. The triangular arrays are
made of 12 columns by 12 rows and have $S\,=\,
160\,\mu$m$^2$. All arrays are underdamped
and typical values of the McCumber parameter\cite{Lih} $\beta_c$
at $T=4.2\,$K are around 200.
The parameter $\beta_L$ characterizing
the influence of self-inductance \cite{FilWis,Barah98} changes from
$\beta_L~\simeq~0.5 $ for triangular arrays to $\beta_L~\simeq~2$
for square arrays.
\begin{figure}[!tbp]
\centering\leavevmode
\mbox{\epsfbox{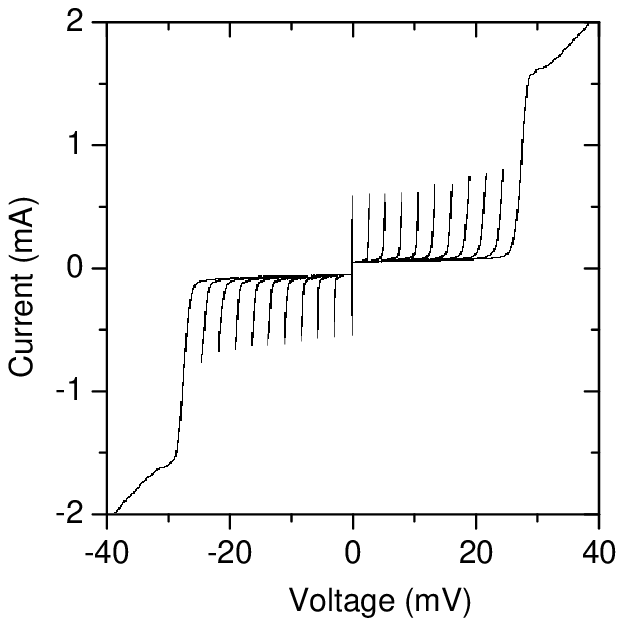}}
\caption{Current-voltage characteristics of a $10\,\times\,10$ array.}
\end{figure}

\noindent The bias current $I$ is injected uniformly via external
resistors. The voltage is read in the direction along the bias, both
across the whole array and across several individual rows.
The $I$-$V$ curves are digitally
stored while sweeping the bias current occurs.
The $I$-$V$
characteristic of the square array is shown in Fig.~2.
For every bias
polarity, the number of branches is equal to the number of rows which
are at the gap voltage state $V_g~\simeq~2.7\,$mV, and the highest gap
voltage corresponds to the sum of gap voltages of the 10 rows. By
choosing a bias point at a certain gap voltage, we can select the
number of rows which will be in the resistive state, while the other rows
remain in the superconducting state.

On the hysteretic part of the $I$-$V$ curves, we have
systematically
observed a fine branching around the row gap voltages (Fig.~3(a)).
The branching
has been detected simply by recording a large number of $I$-$V$ curves
in the absence of
externally applied magnetic field
and at a constant sweep frequency and
temperature ($T\,=\,4.2\,$K). It was always
possible to choose a stable dc bias point on a particular branch.
The fine branching of the $I$-$V$ curves
around the gap voltages was found to be a typical feature of all studied
arrays.

In order to determine the actual distribution of resistive paths in
the array, we used the method of LTSLM \cite{Sivakov}. The LTSLM
uses a focused laser beam for local heating of the sample. The local
heating leads to an additional dissipation in the area of several
micrometers in diameter. The laser beam induced variation of the voltage
drop across the whole sample is recorded versus the beam coordinates.
This method allows to visualize the junctions that are in the
resistive state. In order to increase the sensitivity and spatial
resolution, the intensity of the beam is modulated at a frequency of a
several KHz and the voltage response of the array is detected phase
sensitively by lock-in technique.

\begin{figure}[tbp]
\centering\leavevmode
\mbox{\epsfbox{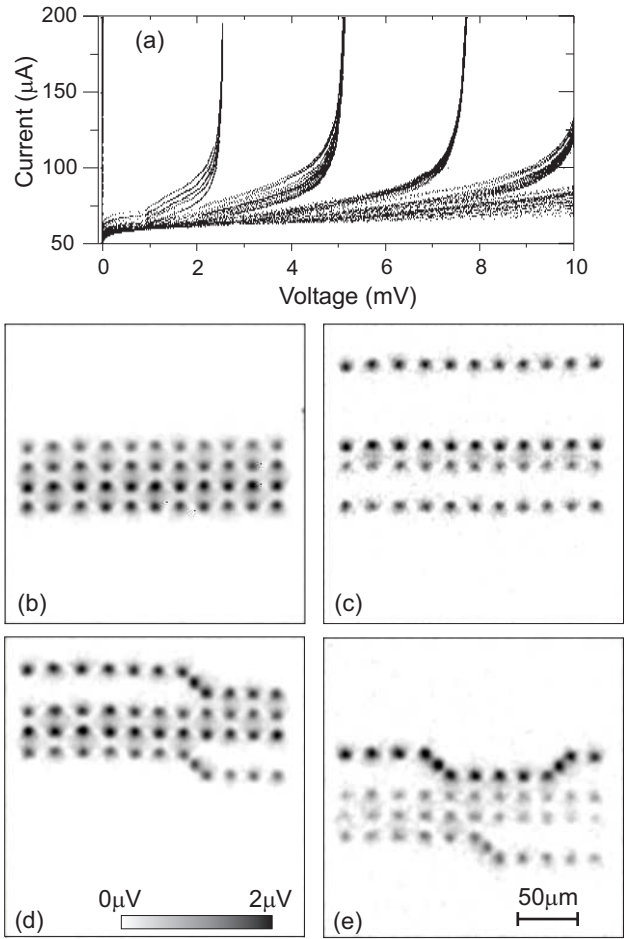}}
\caption{(a)
Enlargement of digitally stored $I$-$V$ curves in the region between zero
and $4V_g$; (b-e) LTSLM images of
resistive states  of the array at the bias points close to the
$4V_g$.}
\end{figure}

By using the LTSLM, we have systematically imaged different resistive
configurations biased at various fine
resistive
branches of the $I$-$V$
curves. The experimental
procedure is the following: the
number $n$ of switched
rows is fixed by biasing the array at voltages
$V\simeq nV_g$. Then LTSLM images of the sample are recorded at a
constant bias value. Typical images from the square array are shown in
Fig.~3(b)-(e).
The black spots correspond to the junctions that are in the resistive
state. The junctions that are in the superconducting state do not
appear on the array image. As expected, the images show that the
number of rows which are in the resistive state is equal to the number
$n$ of gap voltages selected by the bias point.
To trap a
different resistive configuration, we always increased the bias
current $I$ above $I_{c}^{\rm array}$, the array critical current,
and then reduced it to the low level.

Similarly to the experimental results of Ref.~\onlinecite{Lach94}, we have
found
that most of the images show  various combinations of {\it straight}
resistive and superconducting rows (Fig.~3(b),(c)). However, the most striking feature of our measurements
is that the resistive lines are not always straight, but may undergo a
{\it meandering} towards the neighboring row involving one of the
horizontal junctions in the resistive state, see Fig.~3(d).

We have found that such a broken symmetry of row switching systematically
appears in all studied arrays.
Moreover, we have observed that the horizontal
junctions switched to the resistive state were distributed randomly inside
the array and have found no tendency for the meanders to occur at the same
places. Thus, we suppose that the meandering is not predominantly due
to any disorder in the junction parameters. The meandering
of resistive paths is a rare event and its probability does not
exceeds $1\%$ per horizontal junction of single switched row. Therefore, the
observation of deviation of
resistive paths from the straight lines is easier for larger arrays.
We have also observed rare resistive states with two switched horizontal
junctions in one row, as shows Fig.~3(e). Sometimes even more
complicated distributions of switched junctions were registered
[Fig.~4[(a),(b)].
\begin{figure}[!tbp]
\centering\leavevmode
\mbox{\epsfbox{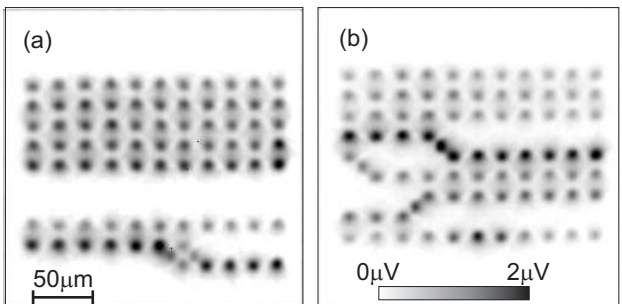}}
\caption{LTSLM images of more complicated dynamic states
observed at the bias points
close to $7V_g$ voltage region.}
\end{figure}
In the following, we present numerical simulations of the dynamics of
underdamped arrays.
Calculations were performed using the resistively shunted model
for Josephson junctions and the usual analysis for superconductive loops
with only self-inductances taken into account, {\it i.e.}
the Nakajima-Sawada equations
\cite{Nak81}.
For
a discussion of these equations and the application limits see
Refs. \cite{FilWis,Barah98}.
The phase configuration of a row switched state was used as initial
condition for the next run.

We simulated large
($10\,\times\,10$) and small
($2\,\times\,10$) square arrays with various parameter $\beta_L$ between
0.5 and 2. In all cases, the simulations well reproduce the
branching of $I$-$V$ curves and
meandering character of the finite voltage paths.
Moreover, in order to check the influence of self-inductance effects the
simulation were carried out for the arrays with a small parameter
$\beta_L~=~0.1$.
A simulated $I$-$V$ curve with three different branches and the
corresponding voltage distributions for the $2\,\times\,10$ array
are shown in Fig.~5.
Resistive states with both single and
double meanders have been trapped (Fig.~5). The appearance of the
meandering is qualitatively similar to the experimental images
shown in Fig.~3(b-e).
The simulations have been carried out in the presence of finite
magnetic field in order to prevent the simultaneous switching of
all the array rows \cite{Barah98}.
It leads to the decrease of the critical current and the appearance of
the flux-flow region where $I$-$V$ curves are highly non linear (Fig.~5).

We have found that, in the absence of disorder, the probability of the
appearance of a meander in two row array is about $ 1.8 \%$ per
horizontal junction. Surprisingly, in the presence of specially
introduced disorder
up to $20 \%$ the numerical simulations show no increase of the rate of
appearance and no preferred position of the meanders.
Due to this
fact, we conclude that the broken symmetry
row switching appears due
to an intrinsic instability
\begin{figure}[!htbp]
\centering\leavevmode
\mbox{\epsfbox{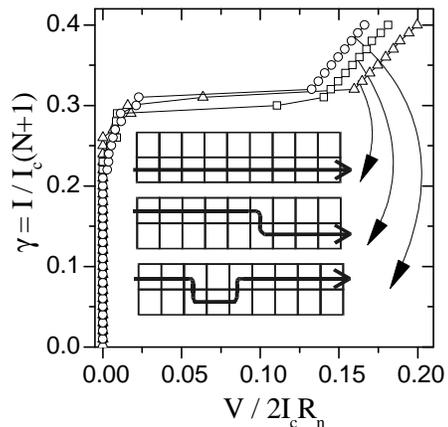}}
\caption{Numerically simulated $I$-$V$ characteristics of a two row array
and the corresponding voltage paths at $\gamma = 0.4$.
$I_c$ is the single junction critical current.}
\end{figure}
of the superconducting state, {\it i.e.}
a change
of the initial conditions can lead to the appearance
of a different percolative voltage path.
The numerical analysis of the two row array also allows to find
out some peculiarities of the new dynamic states.
When the junctions
switched to the
resistive state form a straight line in the top or bottom
row (no meandering), the array has the highest resistance, $R_n/(N+1)$
(triangle symbols in Fig.\,5),
where
$N$ is the total number of cells
in the row and $R_n$ is the normal resistance of one junction.
In the case when the
voltage path with one meandering occurs, the array shows a lower
resistance, $R_n/(N+2)$ (square, Fig.\,5).
Now there is one more
junction
connected in parallel to the
resistive path, with respect to the case of no meandering.
For the same reason,
the meanders across two  horizontal
junctions corresponds, to the
resistance $R_n/(N+3)$ (circle, Fig.\,5).
The presence of meanders shifts the array $I$-$V$ curve to the left and
this resistance scaling
explains the experimentally observed fine branching of the $I$-$V$ curves.

We have found that the above dynamic states are stable in a wide range of
voltages, {\it i.e.} we have not observed direct switching between
branches. This is due to the too large energy necessary for the
junction capacitance recharge in underdamped arrays \cite{Lih}. So,
only switching from the superconducting state allows to trap these dynamic
states.

Our analysis using the Kirchhoff's current laws shows that, in the case
of meandering, the dc superconducting
current flows via the horizontal
junctions. Moreover,
this current increases from the row boundary to
the cell where the meandering occurs.
This current configuration is
different from that without
meanders. In the latter case the
horizontal junctions are not active and dc superconducting current
flows only via the vertical junctions.

\begin{figure}[!htbp]
\centering\leavevmode
\mbox{\epsfbox{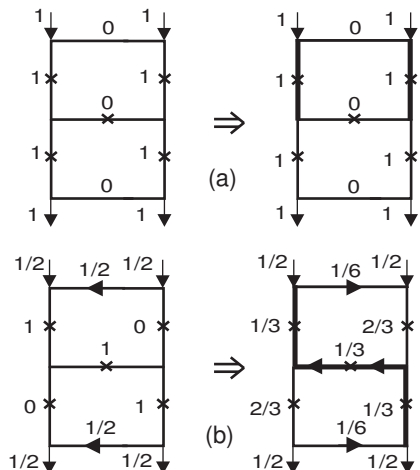}}
\caption{ Sketches of metastable superconducting and
resistive states for a small array with two plaquettes containing
three junctions per cell: (a) mesh currents are equal to zero, (b)
mesh currents are different from zero. Thick solid lines
show the parts of the
array switched to resistive states, and numbers are the local dc
currents normalized to $I_c$. }
\end{figure}
%
All these features can be explained, at least qualitatively, by the
presence of
metastable superconducting states in 2D Josephson arrays.
We analyze these states
for a simple
case of two plaquettes with 5 junctions (Fig.~6).
In the absence of magnetic field, the most stable superconducting
state is the one with all mesh currents equal zero. At the bias
current $I=2I_c$
the vertical junctions switch from superconducting state to
resistive state with straight resistive rows (dc mesh currents are
still zero) (Fig.~6(a)).
However,
this system has another peculiar state with finite
values of mesh currents, as shown in Fig.~6(b). At $I=I_c$ this system can support a superconducting state
with the Josephson phases of the horizontal junction and one vertical
junction per row being equal to $\pi/2$. The Josephson phases of the
remaining vertical junctions are equal to $\pi$. This metastable state
supports finite mesh currents $I_c/2$ and switches to the dynamic
(resistive) state with a meander via the horizontal junction. The mesh
currents decrease to the value of $I_c/6$ and change sign during
the switching process, which results in a stable dynamic state with
meandering. Thus, the dynamic states with
meandered dc voltage drop are "fingerprints" of metastable
superconducting states with non zero mesh currents and active
horizontal junctions.
The state appears only when there are at
least three junctions per cell and it exists also in the limit of zero
linear self-inductance and zero magnetic field \cite{comment2}.
That is
different from the case of single- and double-junction SQUIDs, where
metastable superconducting states appear only in the limit of large
self-inductance \cite{Lih}.
Since the above discussed metastable state exists
in the limit of zero inductance, the small arrays with few plaquettes
can be suitable system to
observe the effect of macroscopic quantum coherence at low
temperatures\cite{Lih,QC}.

It appears that also large 2D arrays can provide
various metastable superconducting states that lead to the broken
symmetry in row switching process. Theoretical and experimental
investigation of these metastable superconducting states and their
dependence on the externally applied magnetic field are in progress.


This work was partially supported by the European Office of
Aerospace Research and Development (EOARD), the Alexander von Humboldt
Stiftung, the German-Israeli Foundation, and the German-Italian
DAAD/Vigoni exchange program.


\begin{references}
\bibitem{comment2} A similar state can be found in a single
anisotropic plaquette with three junctions per cell when the parameter
of anisotropy $\eta=I_{ch}/I_{cv}>1/(2\sqrt{2})$ (M. V. Fistul, to be
published). Here, $I_{ch}$ and $I_{cv}$ are, respectively, the critical
currents of the horizontal and vertical junctions.
\end{references}
\end{document}